\documentclass[aps,prl,twocolumn,showpacs]{revtex4}
\usepackage{graphicx}
\usepackage{amssymb}

\begin{document}

\title{Scaling and precursor motifs in earthquake networks}

\author{Marco Baiesi} 
\email{baiesi@pd.infn.it}
\affiliation{INFM, Dipartimento di Fisica, Universit\`a di Padova, 
I-35131 Padova, Italy.}
\date{\today}
\begin{abstract}

{A measure of the correlation between two earthquakes
is used to link events to their aftershocks, generating a growing
network structure.
In this framework one can quantify whether an aftershock is close or far, 
from main shocks of all magnitudes. 
We find that simple network motifs involving links
to far aftershocks appear frequently before 
the three biggest earthquakes of the last 16 years in Southern California.
Hence, networks could be useful to detect symptoms typically preceding major
events.
}
\end{abstract}

\maketitle

A fundamental open issue in the field of seismicity
is whether earthquakes are to some extent
predictable or not~\cite{scholz_book}. There are conflicting
points of view about this~\cite{scholz_book,nature-debate}.
Nevertheless, phenomenological approaches have been used for some decades
 to formulate algorithms for earthquake 
prediction~\cite{keilis02:_rev,keilis02:_PNAS,zaliapin02:_corr-length,keilis04:_chains}, sometimes based on the search for complex (long-range) 
correlations~\cite{keilis02:_rev}.

Insight into the issue of seismicity and maybe
of earthquake prediction can be 
obtained by measuring the correlations between any pair of earthquakes.
One  method to estimate the amount of correlation was
put forward in Ref.~\cite{baiesi04} (see also~\cite{baiesi04c}), 
based on the statistical properties of earthquakes.
If epicenters are  distributed with a fractal dimension $d_f$,
the mean number of events within 
an area of radius $l$ should scale as $l^{d_f}$.
According to the Gutenberg-Richter law~\cite{scholz_book},
the number of these events with magnitude $\ge m$ is proportional to
$10^{-b m}$, with $b\approx 1$. Of course, the number of these events
is on average also 
proportional to the time $t$ we have been spending to record them.
Hence, globally the mean number of events scales with the size of the
space-time-magnitude window as $ n \simeq K\, t\, 10^{-b\, m}\, l^{d_f}$,
where $K$ is a constant related to the seismic activity.
When a new event $j$ takes place, it defines a point of view
from which one can assess whether past seismic events appear unusual
or usual, with respect to their expected average number.
Indeed, 
any pair of events $(i, j)$, separated by a time interval $t_{ij}$ and a 
distance $l_{ij}$, defines an expected
number of events $n_{ij} = K\, t_{ij}\, 10^{-b\, m_i}\, l_{ij}^{d_f}$,
where $m_i$ is the magnitude of the first event.

One finds small $n_{ij}$ values when $j$ occurs immediately after $i$,
very close to $i$, and if $i$ has a large magnitude.
A very small $n_{ij}$ value means that an event 
with magnitude $m_i$ had very small probability to occur in the space-time 
window defined by event $j$.
Since such a case should rarely take place at random, 
its actual occurrence tells us that $i$ and $j$  are correlated.
Furthermore,
the smaller is $n_{ij}$, the more unusual is event $i$ 
``with respect to $j$'', the more $i$ and $j$ are 
correlated~\cite{note1},
as it was argued in Ref.~\cite{baiesi04}.
Hence, one can adopt $n_{ij}$ as a metric for quantifying correlations between 
events. On the basis of $n_{ij}$ one can also  
build a network of earthquakes~\cite{baiesi04}
by drawing an oriented link to a new event $j$ only from the event $i$
giving the smallest  $n_{ij}$ value (denoted as $n_j^*$).  
In this pair, we call event $i$ the ``main shock'' 
and $j$ is the ``aftershock'' even if $m_j>m_i$~\cite{note_after}.

In this Letter we examine such earthquake correlation graphs
by means of tools of network theory.
We show that the notion of distance at the basis of the network
construction underlies remarkable statistical scaling properties, which
should reflect basic mechanisms of earthquake formation and propagation. 
We also find that some simple motifs (small pieces composed by a few nodes 
and links~\cite{milo02:_motif}) could constitute an 
interesting kind of precursor of major events.
The study of the  motif occurrences
is a strategy to understand the properties of the systems
described by networks~\cite{milo02:_motif}.
For example,
it is currently believed that understanding the statistics of simple motifs
in protein-protein interactions and transcription regulatory 
networks can help to understand the
metabolism~\cite{milo02:_motif,motifs}.

The catalog we have analyzed is maintained by the Southern California (SC)
Earthquake Data Center~\cite{cata1}.
Data in the period ranging from the 1st of January 1984 to the 31st
of December 2003, and
earthquakes with magnitude $m \ge m_< = 3.0$ are considered ($8858$ events). 
In the area covered by the catalog the Gutenberg-Richter law  holds with 
$b\simeq 0.95$~\cite{bak02:_unified}, and $d_f=1.6$~\cite{corral03:_unified}.
Quantities are always measured in MKS units.

We examine the three-dimensional distribution of earthquakes,
taking into account their epicenters (latitude and longitude)
and depths, i.e., their hypocenters. The spatial separation 
between events is given by the Euclidean
distance between their hypocenters, and the fractal dimension 
of hypocenters is supposed to be $D_f = 1 + d_f=2.6$.
The metric we use is then $n_{ij}=K'\,l_{ij}^{D_f} 10^{-b\, m_i} t_{ij}$.
Links reliably denoting correlations have $n_{ij}\le n_c$, with a
suitable threshold $n_c$~\cite{baiesi04,note2}. 
In order to define a selection procedure independent of the constant $K'$, 
here we use $n_c =  \langle n^*\rangle / 10 $, where
$\langle n^*\rangle$ denotes the average of all $n_i^*$
with $i=2,\, 3,\, \ldots, j-1$.

If at most one incoming link per node is allowed,
the network has the form of a growing tree~\cite{baiesi04}. 
We  relax this constraint because we want a richer network structure,
with abundance of motifs like triangles of linked nodes, which are
usually associated with the presence of non-trivial correlations within 
networks~\cite{networks}.
Nearly optimal incoming links to a new event have $n_{ij}$ slightly greater
than their minimum value $n_j^*$ and are the first candidates to be added to 
the tree structure: hence, we choose to draw a link when
$n_{ij}\le n_c$ and $n_{ij} \le \phi\, n_j^*$, with constant 
$\phi> 1$ (this procedure is also suggested by the fact that data 
from catalogs have experimental errors).
We set $\phi=10$, obtaining roughly $2$ outgoing links per node, 
but other similar values do not considerably alter the results.

\begin{figure}[!tb]
\includegraphics[angle=0,width=8.2cm]{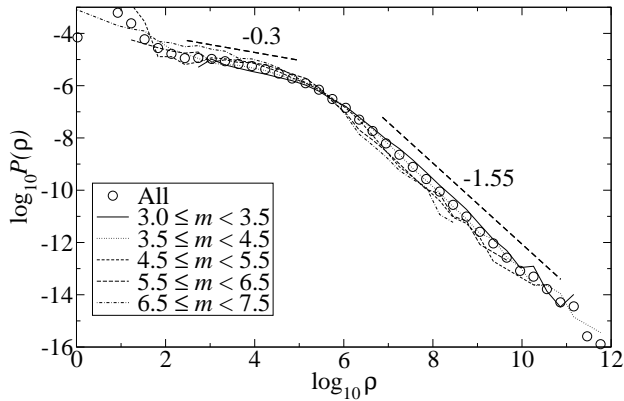}
\caption{Log-log plot 
of the global distribution $P(\rho)$ (circles), and of the
distributions $P_{[m_1,m_2)}(\rho)$ 
generated by earthquakes with magnitude in ranges $[m_1,m_2)$ (see legend). 
Two power-law regimes  
(with relative exponents) are evidenced by dashed straight lines.
\label{fig:PDF}}
\end{figure}

Our analysis of the precursory phenomena 
is based on the statistics of the quantity
\begin{equation}
\rho_{ij} = l_{ij}^{D_f}\, 10^{-b\, m_i} \label{eq:rho}\,,
\end{equation}
which is the space-magnitude part of the metric values $n_{ij}$ associated
with drawn links.
In Fig.~\ref{fig:PDF} we show its distribution $P(\rho)$.
In addition, we also plot the distributions of $\rho$ relative to links 
departing from shocks in ranges of magnitudes $[m_1,m_2)$, 
denoted as $P_{[m_1,m_2)}(\rho)$.
Two distinct power laws appear 
in  $P(\rho)$ as well as in all  $P_{[m_1,m_2)}(\rho)$ considered.
For $\rho\to0$, $P(\rho) \sim \rho^{-\alpha}$, 
with $\alpha\simeq 0.3$. In the regime $\rho\to\infty$ instead 
$P(\rho) \sim \rho^{-\beta}$, with $\beta\simeq 1.55$.
Since all $P_{[m_1,m_2)}(\rho)$ are quite well overlapped,
and the aftershock distances vary weakly with time after an event (not shown),
a length $l_m = \rho^{1/D_f} = 10^{(b/D_f)m}$ is a good unit for measuring
the distance of aftershocks from an event of magnitude $m$.
Thus, the exponent $\sigma = b/D_f\simeq 0.37$ might justify the
rescaling of aftershocks distances with a
factor $10^{\sigma m}$, as it was done in Ref.~\cite{baiesi04}
($\sigma\simeq 0.4$ there).

The distributions $P(\rho)$ describes 
a property of individual correlations between pairs of earthquakes,
from which we clearly see that two classes of aftershocks exist, corresponding 
to the two regimes of $P(\rho)$.
A geophysical explanation of these two regimes 
 could be related to the hierarchical fault structure:
possibly, small $\rho$ are connected to the conventional
 aftershocks within the rupture area, while the high $\rho$ region could 
be determined mainly by inter-fault aftershocks, which are also detected by
our method.

\begin{figure*}[!bt]
\includegraphics[angle=0,width=16.8cm]{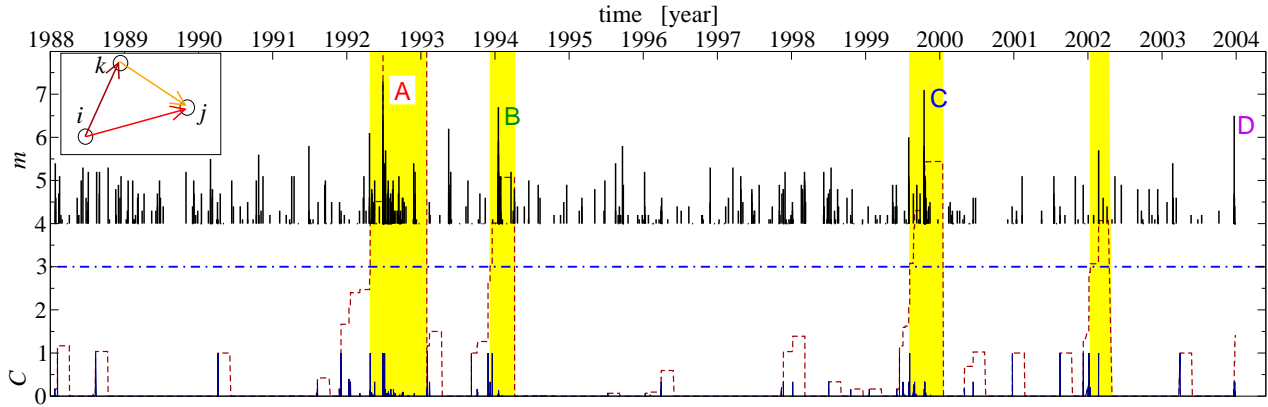}
\caption{(Color online)
Time series of event magnitudes (above, only $m\ge 4$ are shown) and of special
clustering $C$ of events (below). 
Landers (A), Northridge (B), Hector Mine (C), and San Simeon (D) are 
the four biggest events since 1988 in the catalog. 
The integrated signal $C_I$ is shown as a dashed line, while
the horizontal dot-dashed line represents the threshold value $C_H=3$: 
when $C_I>C_H$, alarms are declared (shaded areas, yellow online). 
Inset: sketch of a triangle of linked events, 
which is ``special'' if $\rho_{i k}>\rho_0$.
\label{fig:ts}}
\end{figure*}

A wide area of aftershock activity, as quantified by a large $\rho$ value, 
may be favored by high stresses within the crust, and hence may be 
related to the periods prior to strong earthquakes. During these periods,
it is also reasonable to find complex correlations in the
stress field~\cite{sykes_jaume99:_review}. 
We have tested the possibility that these phenomena are highlighted by
peculiar network motifs, i.e., by studying the
local topological structure of the growing network of earthquakes.
Indications supporting our hypothesis can be found by modifying the notion of
local clustering coefficient of a node, which is normally given by
the fraction of triangles it forms with its 
neighbors~\cite{networks}.
In order to meet our former
requirements, the motifs we study here are special triangles (ST), 
in which the $\rho$ value carried by the first link 
($i$-$k$ link in the Inset of Fig.~\ref{fig:ts}) 
is larger than a given threshold $\rho_0$. 
The special clustering coefficient of a new node $j$ 
is then $C_j = \Delta_j / \Delta_j^{\max}$, where 
$\Delta_j$ is the number of ST 
it forms with its $\kappa_j$ main shocks, 
and $\Delta_j^{\max} = \kappa_j(\kappa_j-1) /2$.
By definition $C_j = 0$ if $\kappa<2$.

\begin{figure}[!bhtp]
\includegraphics[angle=-90,width=7.7cm]{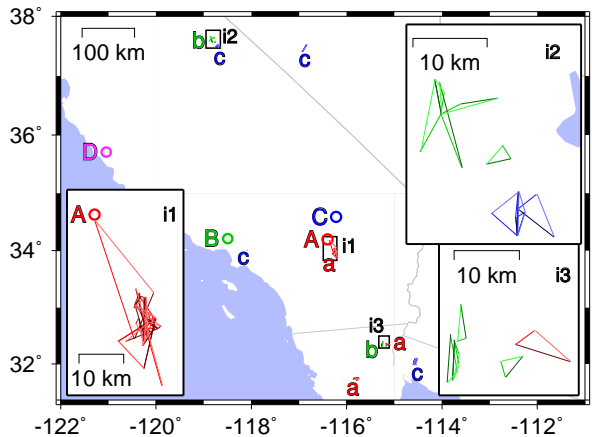}
\caption{(Color online)
Location of big events (circles, same capital letters discussed in
the text and in Fig.~\ref{fig:ts}), 
and of precursor patterns (ST), marked with the same letter (and color online) 
of the relative
big shock. The three insets are enlargements of areas with ST.
Color tones of the three links in a triangle follow the same order as in the 
Inset of Fig.~\ref{fig:ts}; in particular the older link is darker. 
\label{fig:geo}}
\end{figure}

\begin{figure}[!bhtp]
\includegraphics[angle=0,width=7.7cm]{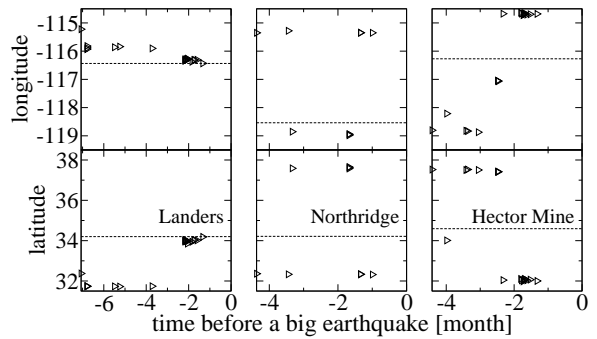}
\caption{Longitude and latitude of the last node of ST,
during the period when $C_I>0$ before Landers, 
Northridge and Hector Mine events.
The coordinates of the big events are plotted as dashed lines.
There is a clear convergence of the ST to
the Landers epicenter [see also Fig~\ref{fig:geo}(i1)]. 
\label{fig:seq}}
\end{figure}

To show that ST may be precursors of strong events we
proceed as follows:
The first three years of the catalog are used to obtain an initial estimate
of $\langle n^* \rangle$.
During the next year we just add links, to avoid possible problems
arising from the analysis of a network where links to old events are lacking.
Then, from the beginning of 1988,
an algorithm analyzes the signal given by the
$C$ value, evaluated for each event when it takes place. 
When $C>0$, we start an integration of the $C$ signal, called
$C_I$, which is reset to zero if $C=0$ for a period $T_0$. 
Values $T_0=60$ days and $\rho_0=10^7$ yield a reasonable overall
rate of $C>0$ values (spikes $0<C\le1$ in Fig.~\ref{fig:ts}),
avoiding the saturation of $C_I$, which is the signal that we think 
is somewhat proportional to the seismic hazard in the region.
The periods when $C_I$ is greater than a constant threshold
$C_H=3$ are declared as alarms.

Figure~\ref{fig:ts} suggests that there is a relation between
alarm times and the occurrence of the three biggest events in the catalog: 
for Landers event [$m=7.3$, labeled with (A)], 
alarm would have started $9$ weeks before its occurrence,
for Northridge [(B), $m=6.7$] one had to wait $6$ weeks
 after the declaration of the alarm, while the alarm before
Hector Mine [(C), $m=7.1$] started $10$ weeks in advance. 
Thus, they would have been predicted in the short term.
The San Simeon event [(D), $m=6.5$] instead was not within an alarm time,
while an alarm was also 
declared in a period when the biggest event had $m=5.7$.

The spatial location of the precursor motifs is another interesting issue.
Fig.~\ref{fig:geo} and Fig.~\ref{fig:seq}
show the distribution of ST giving rise to the alarms (i.e.~when $C_I>0$)
before the three biggest events.
In Fig.~\ref{fig:geo}, small letters corresponding to the big event
ones denote areas with ST, 
and three insets show enlargements of some of them.
Excluding a cluster of ST which would have indicated the future 
location  of Landers epicenter 
[Fig.~\ref{fig:geo}(i1)], ST do not appear close to the location of the  
incoming big events, in agreement with the idea 
that the preparation of an earthquake is not localized around its future 
source (see~\cite{keilis02:_rev} and references therein).

A plausible explanation of both this delocalization of the precursor 
patterns with respect to the big shock and the relation between high 
$\rho$ values and and strong earthquakes might come from the critical
point scenario~\cite{sykes_jaume99:_review,critical_point},
in which  a big event represents a finite time 
singularity~\cite{sammis-sornette02:_PNAS}. 
Indeed, as in the theory of critical phenomena, 
a suitably defined correlation length shows a singular behavior 
diverging prior to big 
earthquakes~\cite{zoeller01:_corr-length,zaliapin02:_corr-length}. 
This length is evaluated by a procedure which 
sums the distances between events which are not aftershocks.
Due to our results,
we believe that aftershock distances may be a complementary indicator
of long range correlations, and in particular that relatively 
far aftershocks could be a typical symptom of
 an incoming strong earthquake.
Notice that we obtain useful informations  also from the statistics of
the aftershocks of the numerous minor earthquakes,
in agreement with the idea that the latter are active 
players in seismicity~\cite{helms02:_trig}.

\begin{figure}[!tb]
\includegraphics[angle=0,width=7.7cm]{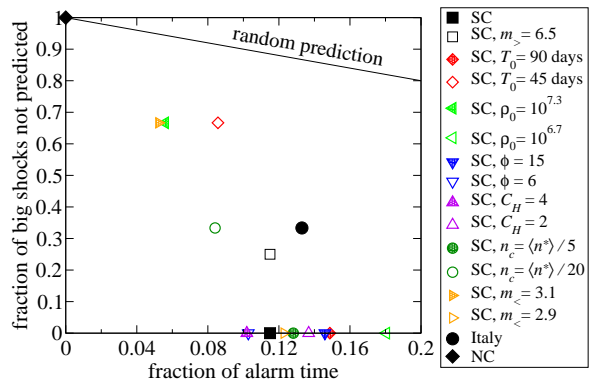}
\caption{(Color online) Error diagram. Symbols are associated 
with geographic zones and eventually with a modified parameter (see text).
The line represents the  performance of a random alarm declaration.
\label{fig:Molchan}}
\end{figure}

To assess the stability of our simple algorithm, 
in Fig.~\ref{fig:Molchan}
we have plotted an error diagram~\cite{molchan97}
where the fraction of events with $m\ge m_>$
that are not predicted is shown as a function of the fraction of alarm time. 
In the diagram, the performance of a random alarm declaration is represented
by a line joining the point $(0,1)$ with $(1,0)$. 
Starting from the point ($n_c=\langle n^* \rangle/10$, 
$\phi=10$, $\rho_0=10^7$, $C_H=3$, $T_0=60$ days, $m_< =3.0$, $m_>=6.7$) 
in the parameter space, we have varied one of
the parameters per time, around its initial point, and plotted the relative
performance in Fig.~\ref{fig:Molchan}.
One clearly see that the algorithm does better than a random alarm 
declaration, and that it is reasonably stable. 

The case illustrated in this paper shows that a translation of issues of
seismicity into a network problem can be a fruitful approach.
In order to have further insight on this possibility, we have analyzed two 
other catalogs, centered around Northern California (NC) and 
Italy~\cite{cata2}, and covering the same time span of our SC catalog. 
We have used the same
parameters of SC, but for NC we set $m_> = 6.5$ to include both S.~Simeon
and Loma Prieta (1989, $m=7$) events in the big shock list.
The algorithm does not recognize any of the two NC big events (no alarms 
declared, see Fig.~\ref{fig:Molchan}).
In Italy we set $\rho_0 = 10^8$ and a shift of the magnitudes
($m_< = 2.5$, $m_> = 5.8$) is necessary in order to
include the two largest events
(Umbria 1997, $m=6$ and Molise 2002, $m=5.9$) in the big shock list
and a considerable number of smaller ones in the analysis. 
In this case, $4/6$ of the big events are predicted, including the two
most disruptive ones, with a fraction of alarm time $\approx 0.13$, as shown
in  Fig.~\ref{fig:Molchan}.

In summary, by means of an appropriate metric quantifying
the amount of correlation between earthquakes, 
aftershocks of any event can be identified.
Aftershock distances from a shock of magnitude $m$ are properly measured by a
length unit scaling as $10^{0.37 m}$.
This information  has been combined with a 
study of the local topology of the growing network of earthquakes, 
to show that simple motifs embodying links to unusually
far aftershocks appeared frequently
before Landers, Northridge and Hector Mine events in Southern California.

The author thanks A.~Kabak\c c\i o\u glu, E.~Orlandini, M.~Paczuski and 
 A.~L.~Stella for the useful comments and discussions.
Support from INFM-PAIS02 is acknowledged.


\begin{thebibliography}{27}
\bibitem{note1} This way of thinking is rather general and can be applied
to contexts different from seismicity as well.
\bibitem{note_after}
As discussed in~\cite{baiesi04},  
the standard aftershock collection instead would count all events following
a stronger earthquake, in a given time window and far less than a maximum 
distance.
\bibitem{note2} In order to avoid too long links bewteen nearly simultaneous
far events, a minimum time interval $t_{\min} = 60$ s between two events is
imposed. In the metric, a minimum distance $l_{\min} = 100$ m is also used.
\end{thebibliography}

\end{document}